\documentstyle[aps,epsf]{revtex}

\begin{document}

\title{Bianchi Type I Universes with dilaton and magnetic fields}
\author{T. Harko\footnote{E-mail: tcharko@hkusua.hku.hk} and M. K. Mak\footnote{E-mail:mkmak@vtc.edu.hk}}
\address{Department of Physics, The University of Hong Kong,
Pokfulam Road, Hong Kong, P. R. China.}
\maketitle

\begin{abstract}

We consider the dynamics of a Bianchi type I space-time in the presence of
dilaton and magnetic fields. The general solution of the Einstein-dilaton-Maxwell
field equations can be obtained in an exact parametric form. Depending on the numerical values of
the parameters of the model there are three distinct classes of solutions. The time evolution of the mean anisotropy, shear and deceleration parameters
is considered in detail and it is shown that a magnetic-dilaton anisotropic Bianchi type I geometry does not isotropize, the initial
anisotropy being present in the universe for all times.
\end{abstract}

\section{Introduction}

In a series of recent papers \cite{SaSa98}, \cite{BaGh99}, Bianchi
type I cosmological models based on the field equations derived from the
action 
\begin{equation}\label{1}
S=-\frac{1}{2}\int d^{4}x\sqrt{-g}\left[ R-2\gamma \left( \nabla \phi
\right) ^{2}+e^{-2\alpha \phi }F^{2}\right],
\end{equation}
where $\alpha =const.>0$, $\gamma =const.>0$ have been considered. In Eq. (\ref{1}) $\phi $ is the dilaton field and $F$
is the Maxwell two-form field, with components $F_{\mu \nu }=\partial _{\mu
}A_{\nu }-\partial _{\nu }A_{\mu }$.
The action (\ref{1}) can be obtained from the string frame low energy effective action
\begin{equation}
\hat{S}=-\int d^{4}x\sqrt{-\hat{g}}e^{-2\alpha \phi }\left[ \hat{R}-2\gamma
\left( \hat{\nabla}\phi \right) ^{2}+F^{2}\right] ,
\end{equation}
by means of the conformal transformation $\hat{g}_{\mu \nu }=e^{2\alpha \phi
}g_{\mu \nu }$. As a particular case these models
include Einstein-Maxwell ($\alpha =\gamma =0$) and low energy string theory
($\alpha =2$, $\gamma =2$ ). In order to study the corresponding
anisotropic cosmological model Salim, Sautu and Martins \cite{SaSa98} wrote the field
equations as an autonomous system of differential equations, with the time
coordinate redefined in term of the comoving volume element. Several
particular solutions of Kasner and exponential type have been considered and
the exact general solution for the model has been obtained. The main
conclusions of the authors of \cite{SaSa98} are that the typical
oscillatory dynamics of the model based on the Einstein-Maxwell theory in a
Bianchi type I geometry is not present when approaching the initial
singularity, due to the presence of the dilaton field and that the process
of isotropization is completely absent. But as mentionned by Bannerjee and
Ghosh \cite{BaGh99}, these solutions apparently do not satisfy all the
field equations, the substitution of the solutions into the equation of
motion of the dilaton field leading to some inconsistencies. In their
investigation of the Bianchi type I cosmology Bannerjee and Ghosh \cite{BaGh99} obtained a
simple bi-axial (two equal scale factors) solution, with scale factors
starting from an initial pointlike singularity of zero volume. Hence a
general solution of the Bianchi type I Einstein-Maxwell-dilaton model based
on the action (\ref{1}) seems to be still missing.

The investigation of cosmological models with magnetic fields is definitely
of major physical interest. The possibility of the existence of a
homogeneous intergalactic field (possible of primordial origin) is not ruled
out by observations. In an investigation of a magnetic Bianchi type I
Universe Madsen \cite{Ma89} obtained a bound of the cosmological
magnetic field given by $B\leq 3\times 10^{-4}h\sqrt{\delta \Omega }\left(
1+z_{d}\right) ^{-\frac{1}{2}}G$, where $\Omega =\frac{8\pi G\rho }{3H^{2}}$
is the cosmological density parameter, $h$ is the Hubble parameter measured
in units of $100kms^{-1\text{ }}$, $z_{d}$ is the redshift at which the
anisotropy begins to grow and $\delta =10^{-4}$. Although no magnetism has
been seen yet on a cosmic scale or on a void's scale, magnetism has been
detected in some superclusters of galaxies, clusters of galaxies, halos of
elliptical galaxies, in the disk of spiral galaxies etc. \cite{Va90}. 

From a theoretical point of view Einstein-Maxwell type magnetic models for
the class of hypersurface-orthogonal Bianchi I cosmologies with a $\gamma $
- law perfect fluid and a pure, homogeneous source-free magnetic field have
been recently analyzed by using methods from the qualitative theory if
dynamical systems in \cite{Bl97}. The  main physical results of this investigation are that
if $0<\gamma \le \frac{4}{3}$, magnetic fields do not prevent isotropization in the absence of spatial
curvature. Even in the absence of anisotropic spatial curvature mixmaster-like oscillations also occur.

On the other hand in an attempt to address the potential inherited from string theory to eliminate
the initial cosmological singularity, Gasperini and Veneziano \cite{GaVe93} initiated a program
known as the pre-big bang scenario. The field equations of the pre-big bang cosmology are based on the
low energy effective action resulting from string theory. Pre-big bang cosmological models,
in which there is no need to introduce inflation or to fine-tune potentials, have many attractive features.
Inflation is natural, thanks to the duality symmetries of string cosmology and the
initial condition problem is decoupled from the singularity problem. Finally, quantum
instabilities (pair creation) is able to heat up an initially cold universe and generate a standard hot big bang
with the additional features of homogeneity, flatness and isotropy (for a recent and extensive review of string cosmology see \cite{LiWaCo00}).
Exact anisotropic Bianchi type I cosmological solutions of the low energy string effective action in the presence of homogeneous magnetic fields
have been obtained by Giovannini \cite{Gi99}, \cite{Gi00}. The obtained cosmologies are fully anisotropic
and only quadratic curvature corrections are able to isotropize the geometry of the Universe. The presence
of magnetic seeds can induce further anisotropies, which can, in principle, be present also at late times. 

It is the purpose of the present paper to consider the general solution of
the Einstein-dilaton-Maxwell field equations derived from the action
(\ref{1}). The mathematical description of this system can be reduced
to a single second-order differential equation and the general solution of
the system can be represented in an exact parametric form. The behaviour of
the basic cosmological parameters of observational interest like the
anisotropy and deceleration parameters is also considered in detail. As a general result of our study we show that the dilaton-magnetic Bianchi type I geometry
does not isotropize and hence it is very unlikely that the present day cosmological magnetic fields could be
remnants from a pre-big bang era.

Throughout this paper we use the Landau-Lifshitz conventions \cite{LaLi75}. The present paper is organized as follows. The field equations are written
down in Section II. The general solution of the field equations is obtained
in Section III. In Section IV we discuss and conclude our results.

\section{Field equations, geometry and consequences}

The field equations obtained from the variation of the action (\ref{1}) are
given by: 
\begin{equation}\label{2}
\left( e^{-2\alpha \phi }F^{\mu \nu }\right) _{;\mu }=0,  
\end{equation}
\begin{equation}\label{3}
\phi _{;\mu }^{;\mu }+\frac{\alpha }{2\gamma }e^{-2\alpha \phi }F^{2}=0,
\end{equation}
\begin{equation}\label{4}
R_{\mu \nu }=2\gamma \phi _{;\mu }\phi _{;\nu }+2e^{-2\alpha \phi }\left(
F_{\mu \lambda }F_{\nu }^{\lambda }-\frac{1}{4}g_{\mu \nu }F^{2}\right),
\end{equation}
\begin{equation}\label{5}
\epsilon ^{\alpha \lambda \mu \nu }\partial _{\lambda }F_{\mu \nu }=0.
\end{equation}

The last equation is the Bianchi identity for the Maxwell field. Throughout
this paper we consider natural units with $c=8\pi G=1$. A semi-colon $;$ denotes
the covariant derivative with respect to the metric.

The line element of the Bianchi type I geometry, generalizing to the
anisotropic case the flat Robertson-Walker space-time is given by 
\begin{equation}\label{6}
ds^{2}=dt^{2}-a_{1}^{2}(t)dx^{2}-a_{2}^{2}(t)dy^{2}-a_{3}^{2}(t)dz^{2}.
\end{equation}

For this metric we introduce the following variables \cite{Gr90}: the
volume scale factor $V=\Pi _{i=1}^{3}a_{i}$, the directional Hubble factors 
$H_{i}=\frac{\dot{a}_{i}}{a_{i}},i=1,2,3$, the mean Hubble factor $H=\frac{1%
}{3}\sum_{i=1}^{3}H_{i}$ and we denote $\Delta H_{i}=H-H_{i},i=1,2,3$. From
these definitions we immediately obtain $H=\frac{1}{3}\frac{\dot{V}}{V}$.

Without any loss of generality we assume that the magnetic field is aligned
along the $x$ - direction. Its unique component is $F_{23}=F_{0}=const.$ and
it satisfies both Maxwell equations (\ref{2}) and (\ref{5}). Then $%
F^{2}=F_{\mu \nu }F^{\mu \nu }=2\frac{F_{0}^{2}}{V^{2}}a_{1}^{2}$.

We also introduce the symbol $\varepsilon _{i},i=1,2,3$ defined by $%
\varepsilon _{1}=-1,\varepsilon _{2}=\varepsilon _{3}=+1$. $\varepsilon
_{i} $ has the obvious property $\sum_{i=1}^{3}\varepsilon _{i}=+1$.

With the use of the variables introduced above the non-trivial field
equations (\ref{2})-(\ref{5}) are given by 
\begin{equation}\label{7}
3\dot{H}+\sum_{i=1}^{3}H_{i}^{2}=-2\gamma \dot{\phi}^{2}-F_{0}^{2}\frac{%
a_{1}^{2}}{V^{2}}e^{-2\alpha \phi }, 
\end{equation}
\begin{equation}\label{8}
\frac{1}{V}\frac{d}{dt}\left( VH_{i}\right) =\varepsilon _{i}F_{0}^{2}\frac{%
a_{1}^{2}}{V^{2}}e^{-2\alpha \phi },i=1,2,3,  
\end{equation}
\begin{equation}\label{9}
\frac{1}{V}\frac{d}{dt}\left( V\dot{\phi}\right) =\frac{\alpha }{2\gamma }%
F_{0}^{2}\frac{a_{1}^{2}}{V^{2}}e^{-2\alpha \phi }.  
\end{equation}

Adding Eqs.(\ref{8}) one obtain 
\begin{equation}\label{10}
\frac{1}{V}\frac{d}{dt}\left( 3VH\right) =F_{0}^{2}\frac{a_{1}^{2}}{V^{2}}%
e^{-2\alpha \phi }.  
\end{equation}

Substituting the term $\frac{F^{2}}{2}e^{-2\alpha \phi }$ expressed from Eq.(\ref{10}) into Eq.(\ref{9}) we find 
\begin{equation}\label{11}
\frac{1}{V}\frac{d}{dt}\left( V\dot{\phi}\right) =\frac{\alpha }{2\gamma }%
\frac{1}{V}\frac{d}{dt}\left( 3VH\right),  
\end{equation}
or, equivalently, 
\begin{equation}\label{12}
\dot{\phi}=\frac{\alpha }{2\gamma }\frac{\dot{V}}{V}+\frac{\alpha }{2\gamma }%
\frac{m}{V},  
\end{equation}
with $m$ an arbitrary constant of integration. Integrating Eq.(\ref{12}) we
obtain for the dilaton field 
\begin{equation}\label{13}
\phi =\phi _{0}+\frac{\alpha }{2\gamma }\ln V+\frac{\alpha m}{2\gamma }\int 
\frac{dt}{V},  
\end{equation}
where $\phi _{0}$ is a constant of integration.

We combine now the dilaton field Eq.(\ref{9}) and the gravitational field
equations (\ref{8}) to find 
\begin{equation}\label{14}
\frac{1}{V}\frac{d}{dt}\left( VH_{i}\right) =\varepsilon _{i}\frac{2\gamma }{%
\alpha }\frac{1}{V}\frac{d}{dt}\left( V\dot{\phi}\right) ,i=1,2,3,
\end{equation}
which gives 
\begin{equation}\label{15}
H_{i}=\varepsilon _{i}\frac{2\gamma }{\alpha }\dot{\phi}+\frac{K_{i}}{V}%
,i=1,2,3.  
\end{equation}

$K_{i},i=1,2,3$ are arbitrary constants of integration. With the use of Eq.
(\ref{12}) we can express Eq.(\ref{15}) in the equivalent form
\begin{equation}\label{16}
H_{i}=\varepsilon _{i}\frac{\dot{V}}{V}+\frac{K_{i}+m\varepsilon _{i}}{V}%
,i=1,2,3. 
\end{equation}

From Eqs.(\ref{16}) it follows that the integration constants $K_{i}$ must
satisfy the consistency condition 
\begin{equation}\label{17}
m+\sum_{i=1}^{3}K_{i}=0. 
\end{equation}

The scale factors are obtained by integrating Eq. (\ref{16}) and are given
by 
\begin{equation}\label{18}
a_{i}=a_{i0}V^{\varepsilon _{i}}e^{\left( K_{i}+m\varepsilon _{i}\right)
\int \frac{dt}{V}},i=1,2,3,  
\end{equation}
with $a_{i0},i=1,2,3$ non-negative constants of integration.

We substitute now the dilaton field Eq.(\ref{12}), $H_{i},i=1,2,3$ given by
Eqs.(\ref{16}) and the magnetic field from Eq. (\ref{10}) into the field
equation Eq.(\ref{7}) to obtain the basic equation describing the evolution
of the dilaton and magnetic field filled Bianchi type I space-time: 
\begin{equation}\label{19}
\frac{\ddot{V}}{V}+C\frac{\dot{V}^{2}}{V^{2}}+B\frac{\dot{V}}{V^{2}}+\frac{K%
}{V^{2}}=0,  
\end{equation}
where $C=1+\frac{\alpha ^{2}}{4\gamma }$, $B=\sum_{i=1}^{3}\left(
\varepsilon _{i}K_{i}+m\right) +\frac{\alpha ^{2}}{2\gamma }m$ and $%
K=\frac{\sum_{i=1}^{3}\left( K_{i}+m\varepsilon _{i}\right) ^{2}}{2}+\frac{\alpha ^{2}%
}{4\gamma }m^{2}$.

Denoting $\dot{V}=u$, Eq.(\ref{19}) is transformed into
\begin{equation}\label{21}
\frac{dV}{V}=-\frac{udu}{Cu^{2}+Bu+K}.  
\end{equation}

The general solution of Eq.(\ref{21}) can be formally represented in the
form 
\begin{equation}\label{22}
V=V_{0}e^{-\int \frac{udu}{Cu^{2}+Bu+K}},
\end{equation}
with $V_{0}>0$  a constant of integration.

The physical quantities of observational interest in cosmology are the
expansion scalar $\theta =3H$, the mean anisotropy parameter $A$, the
shear scalar $\Sigma ^{2}$ and the deceleration parameter $q$ defined according
to
\begin{equation}\label{23}
A=\frac{1}{3}\sum_{i=1}^{3}\left( \frac{\Delta H_{i}}{H}\right) ^{2}, 
\end{equation}
\begin{equation}\label{24}
\Sigma ^{2}=\frac{1}{2}\left( \sum_{i=1}^{3}H_{i}^{2}-3H^{2}\right)=\frac{3}{2}AH^2,  
\end{equation}
\begin{equation}\label{25}
q=\frac{d}{dt}\frac{1}{H}-1. 
\end{equation}

With the use of Eqs.(\ref{16}) we can express the mean anisotropy and the
deceleration parameter in the alternative form
\begin{equation}\label{26}
A=8+\frac{2\sum_{i=1}^{3}\left( \varepsilon _{i}K_{i}+m\right) }{VH}+\frac{1%
}{3}\frac{\sum_{i=1}^{3}\left( K_{i}+m\varepsilon _{i}\right) ^{2}}{%
V^{2}H^{2}},  
\end{equation}
\begin{equation}\label{27}
\Sigma ^{2}=H^{2}\left( 12+\frac{3\sum_{i=1}^{3}\left( \varepsilon
_{i}K_{i}+m\right) }{VH}+\frac{1}{2}\frac{\sum_{i=1}^{3}\left(
K_{i}+m\varepsilon _{i}\right) ^{2}}{V^{2}H^{2}}\right). 
\end{equation}

The sign of the deceleration parameter indicates whether the
cosmological model inflates. The positive sign corresponds to standard decelerating models while a negative sign indicates inflation.
The deceleration parameter can be expressed as a function of the variable $u$
in the general form
\begin{equation}
q=2+\frac{3\left( Cu^{2}+Bu+K\right) }{u^{2}}. 
\end{equation}

In the isotropic limit the mean anisotropy parameter is zero.

\section{General solution of the field equations}

In the previous Section we have obtained the basic equations describing the
dynamics of an anisotropic Bianchi type I universe filled with dilaton and
magnetic fields. The field equations of the model can be reduced to a single
equation (\ref{19}) describing the evolution of the anisotropic universe. According to
the sign of the parameter $\Delta =B^{2}-4CK$ the second order differential equation Eq.(\ref{19}) has three distinct
classes of solutions.

For $\Delta =B^{2}-4CK>0$ the general solution of Eq. (\ref{19}) is
\begin{equation}\label{28}
V=V_{0}\left( u-u_{+}\right) ^{m_{+}}\left( u-u_{-}\right) ^{m_{-}},
\end{equation}
where we denoted $u_{\pm }=\frac{-B\pm \sqrt{\Delta }}{2C}$ and $m_{\pm
}=\mp \frac{u_{\pm }}{\sqrt{\Delta }}$. Therefore the general solution of
the field equations can be expressed in the following parametric form, with $%
u$ taken as parameter:
\begin{equation}\label{29}
t-t_{0}=-\frac{V_{0}}{C}\int \left( u-u_{+}\right) ^{m_{+}-1}\left(
u-u_{-}\right) ^{m_{-}-1}du, 
\end{equation}
\begin{equation}\label{30}
H(u)=\frac{1}{3V_{0}}\frac{u}{\left( u-u_{+}\right) ^{m_{+}}\left(
u-u_{-}\right) ^{m_{-}}}, 
\end{equation}
\begin{equation}\label{31}
a_{i}(u)=a_{i0}\left( u-u_{+}\right) ^{p_{i}^{+}}\left( u-u_{-}\right)
^{p_{i}^{-}},i=1,2,3, 
\end{equation}
\begin{equation}\label{32}
A(u)=8+\frac{6\sum_{i=1}^{3}\left( \varepsilon _{i}K_{i}+m\right) }{u}+\frac{%
3\sum_{i=1}^{3}\left( K_{i}+m\varepsilon _{i}\right) ^{2}}{u^{2}},
\end{equation}
\begin{equation}\label{33}
\Sigma ^{2}(u)=\frac{1}{9V_{0}^{2}}\frac{u^{2}}{\left( u-u_{+}\right)
^{2m_{+}}\left( u-u_{-}\right) ^{2m_{-}}}\left( 12+\frac{9\sum_{i=1}^{3}%
\left( \varepsilon _{i}K_{i}+m\right) }{u}+\frac{9}{2}\frac{%
\sum_{i=1}^{3}\left( K_{i}+m\varepsilon _{i}\right) ^{2}}{u^{2}}\right), 
\end{equation}
\begin{equation}\label{34}
q(u)=2+\frac{3C\left( u-u_{+}\right) \left( u-u_{-}\right) }{u^{2}}, 
\end{equation}
\begin{equation}\label{35}
\phi (u)=\phi _{0}+\frac{\alpha }{2\gamma }\ln \left[ \left( u-u_{+}\right)
^{n_{+}}\left( u-u_{-}\right) ^{n_{-}}\right],
\end{equation}
where we denoted $p_{i}^{\pm }=\varepsilon _{i}m_{\pm }\mp \frac{%
K_{i}+m\varepsilon _{i}}{\sqrt{\Delta }},$ $i=1,2,3$ and $n_{\pm }=m_{\pm
}\mp \frac{m}{\sqrt{\Delta }}$. For $u_{\pm }>0$ the condition of the reality of the
physical parameters implies that this  solution is defined only for values
of the parameter $u$ so that $u>u_{+},u_{-}$. For $u_{\pm }<0$ there are no restrictions on the allowed
range of the parameter $u$.

For $\Delta =B^{2}-4CK$ $=0$ the general solution of the field equations is
given by
\begin{equation}\label{36}
t-t_{0}=-\frac{V_{0}}{C}\int \frac{e^{\frac{m_{0}}{u-u_{0}}}}{\left(
u-u_{0}\right) ^{\frac{1}{C}+2}}du, 
\end{equation}
\begin{equation}\label{37}
V(u)=V_{0}\left( u-u_{0}\right) ^{-\frac{1}{C}}e^{\frac{m_{0}}{u-u_{0}}}, 
\end{equation}
\begin{equation}\label{38}
H(u)=\frac{1}{3V_{0}}u\left( u-u_{0}\right) ^{\frac{1}{C}}e^{-\frac{m_{0}}{%
u-u_{0}}},
\end{equation}
\begin{equation}\label{39}
a_{i}(u)=a_{i0}\left( u-u_{0}\right) ^{-\frac{\varepsilon _{i}}{C}}e^{\frac{1%
}{C}\frac{\left( m+Cm_{0}\right) \varepsilon _{i}+K_{i}}{u-u_{0}}},i=1,2,3, 
\end{equation}
\begin{equation}\label{40}
A(u)=8+\frac{6\sum_{i=1}^{3}\left( \varepsilon _{i}K_{i}+m\right) }{u}+\frac{%
3\sum_{i=1}^{3}\left( K_{i}+m\varepsilon _{i}\right) ^{2}}{u^{2}},
\end{equation}
\begin{equation}\label{41}
\Sigma ^{2}(u)=\frac{1}{9V_{0}^{2}}u^{2}\left( u-u_{0}\right) ^{\frac{2}{C}%
}e^{-\frac{2m_{0}}{u-u_{0}}}\left( 12+\frac{9\sum_{i=1}^{3}\left(
\varepsilon _{i}K_{i}+m\right) }{u}+\frac{9}{2}\frac{\sum_{i=1}^{3}\left(
K_{i}+m\varepsilon _{i}\right) ^{2}}{u^{2}}\right),
\end{equation}
\begin{equation}\label{42}
q(u)=2+\frac{3C\left( u-u_{0}\right) ^{2}}{u^{2}}, 
\end{equation}
\begin{equation}\label{43}
\phi (u)=\phi _{0}+\frac{\alpha }{2\gamma C}\left[ \frac{u_{0}+m}{u-u_{0}}%
-\ln \left( u-u_{0}\right) \right],
\end{equation}
where we denoted $u_{0}=-\frac{B}{2C}$ and $m_{0}=\frac{u_{0}}{C}$. In
order to have well-defined (real) physical quantities it is necessary that
the parameter $u$ satisfies the condition $u>u_{0}$.

For $\Delta =B^{2}-4CK$ $<0$ the general solution of the field equations is
\begin{equation}\label{44}
t-t_{0}=-\frac{V_{0}}{C}\int \frac{\exp \left[ \frac{B}{2C^{2}\Delta _{0}}%
\arctan \left( \frac{u+\frac{B}{2C}}{\Delta _{0}}\right) \right] }{\left[
\left( u+\frac{B}{2C}\right) ^{2}+\Delta _{0}^{2}\right] ^{1+\frac{1}{2C}}}%
du,
\end{equation}
\begin{equation}\label{45}
V(u)=V_{0}\left[ \left( u+\frac{B}{2C}\right) ^{2}+\Delta _{0}^{2}\right] ^{-%
\frac{1}{2C}}\exp \left[ \frac{B}{2C^{2}\Delta _{0}}\arctan \left( \frac{u+%
\frac{B}{2C}}{\Delta _{0}}\right) \right],
\end{equation}
\begin{equation}\label{46}
H(u)=\frac{1}{3V_{0}}u\left[ \left( u+\frac{B}{2C}\right) ^{2}+\Delta
_{0}^{2}\right] ^{\frac{1}{2C}}\exp \left[ -\frac{B}{2C^{2}\Delta _{0}}%
\arctan \left( \frac{u+\frac{B}{2C}}{\Delta _{0}}\right) \right],
\end{equation}
\begin{equation}\label{47}
a_{i}(u)=a_{i0}\frac{\exp \left\{ \frac{1}{C\Delta _{0}}\left[ \varepsilon
_{i}\left( \frac{B}{2C}-m\right) -K_{i}\right] \arctan \left( \frac{u+\frac{B%
}{2C}}{\Delta _{0}}\right) \right\} }{\left[ \left( u+\frac{B}{2C}\right)
^{2}+\Delta _{0}^{2}\right] ^{\frac{\varepsilon _{i}}{2C}}},i=1,2,3,
\end{equation}
\begin{equation}\label{48}
A(u)=8+\frac{6\sum_{i=1}^{3}\left( \varepsilon _{i}K_{i}+m\right) }{u}+\frac{%
3\sum_{i=1}^{3}\left( K_{i}+m\varepsilon _{i}\right) ^{2}}{u^{2}},
\end{equation}
\begin{equation}\label{49}
\Sigma ^{2}(u)=\frac{1}{9V_{0}^{2}}\frac{u^{2}\left[ \left( u+\frac{B}{2C}%
\right) ^{2}+\Delta _{0}^{2}\right] ^{\frac{1}{C}}\left( 12+\frac{%
9\sum_{i=1}^{3}\left( \varepsilon _{i}K_{i}+m\right) }{u}+\frac{9}{2}\frac{%
\sum_{i=1}^{3}\left( K_{i}+m\varepsilon _{i}\right) ^{2}}{u^{2}}\right) }{%
\exp \left[ \frac{B}{C^{2}\Delta _{0}}\arctan \left( \frac{u+\frac{B}{2C}}{%
\Delta _{0}}\right) \right] },
\end{equation}
\begin{equation}\label{50}
q(u)=2+3C\frac{\left( u+\frac{B}{2C}\right) ^{2}+\Delta _{0}^{2}}{u^{2}},
\end{equation}
\begin{equation}\label{51}
\phi (u)=\phi _{0}-\frac{\alpha }{4C\gamma }\ln \left[ \left( u+\frac{B}{2C}%
\right) ^{2}+\Delta _{0}^{2}\right] +\frac{\alpha }{2\gamma C\Delta _{0}}%
\left( \frac{B}{2C}-m\right) \arctan \left( \frac{u+\frac{B}{2C}}{\Delta _{0}%
}\right),  
\end{equation}
where $\Delta _{0}=\frac{\sqrt{-\Delta }}{2C}$.

\section{Discussions and final remarks}

In order to consider the general effects of the magnetic field on the
dynamics and evolution of the Bianchi type I space-time we also present the
solutions corresponding to the pure anisotropic dilatonic universe, without
magnetic field, $F_{0}=0$ . These solutions are not new, they have already
been discussed in \cite{ChHaMa00a} and \cite{ChHaMa00b}. In the absence
of the magnetic field the dilatonic universe is described by 
\begin{equation}\label{52}
V=V_{0}t,H=\frac{1}{3t},a=a_{i0}t^{p_{i}},
\end{equation}
\begin{equation}\label{53}
A=\frac{3K^{2}}{V_{0}^{2}}=const.,q=2=const.,\phi =\phi _{0}\ln t,
\end{equation}
where $\phi _{0}=\phi _{0}^{\prime }\sqrt{\frac{1}{4\gamma }\left( \frac{2}{3%
}-\frac{K^{2}}{V_{0}^{2}}\right) }$, with $\phi _{0}^{\prime }$ an
arbitrary constant of integration. $K_{i},i=1,2,3$ are constants of
integration satisfying the condition $\sum_{i=1}^{3}K_{i}=0$. We also
denoted $K^{2}=\sum_{i=1}^{3}K_{i}^{2}$. The coefficients $p_{i}=\frac{1}{3}%
+\frac{K_{i}}{V_{0}},i=1,2,3$ satisfy the relations $\sum_{i=1}^{3}p_{i}=1$
and $\sum_{i=1}^{3}p_{i}^{2}=\frac{1}{3}+\frac{K^{2}}{V_{0}^{2}}$. The
geometry of the dilaton-field filled universe is of Kasner type. A universe
of this type will never experience a transition to an isotropic phase and
its evolution is non-inflationary for all times.

The variation of the volume scale factor $V$ of the magnetic and dilaton
fields filled Bianchi type I space-time is represented, for all three
models, in Fig. 1.

\begin{figure}[h]
\epsfxsize=10cm
\centerline{\epsffile{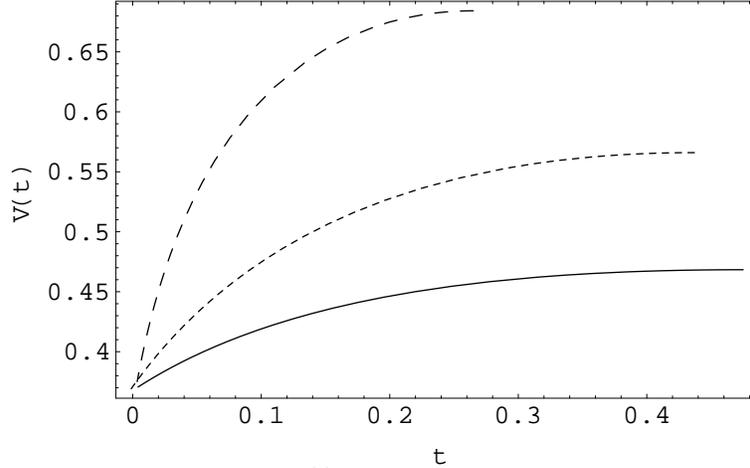}}
\caption{Time evolution of the volume scale factor $V(t)$ of the magnetic-dilaton Bianchi I space-time for different
values of the parameters: $B=4$, $C=1$ and $K=1$ ($\Delta >0$) (solid curve), $B=1$, $C=1$, $m=2.5$ and $K=\frac{1}{4}$ ($\Delta =0$) (dotted curve)
and $B=2$, $C=2$ and $K=2$ ($\Delta <0$) (dashed curve).}
\label{FIG1}
\end{figure}

The evolution of the universe is generally expansionary.
The singularity behaviour depends on the sign of the constants $u_{\pm }$
and $u_{0}$, the real rooots of the equation $Cu^{2}+Bu+K=0$. If all these
roots are positive, then a singular behaviour of the scale factors and of
the volume scale factor cannot be avoided, with the singularity
corresponding to values $u=u_{\pm }$ and $u=u_{0}$ of the parameter. In this case the
range of the parameter $u$ is restricted to $u\geq u_{\pm }$ and $u\geq
u_{0} $. In the case of complex $u_{\pm }$ ($\Delta <0$ ), there is no
singular point in the time evolution of the scale factors $a_{i},i=1,2,3$.
The asymptotic behaviour of the solution in the limit of large $\dot{V}=u$
(that is, for a very rapid expansion of the volume of the universe) can be
obtained from equation (21), by taking the limit $u\rightarrow \infty $.
Therefore we obtain $V\sim u^{-\frac{1}{C}}$ and $V\sim t^{\frac{1}{1+C}}$ ,
respectively. The scale factors are given by $a_{i}=a_{i0}t^{\varepsilon _{i}%
\frac{1}{1+C}}\exp \left[ \frac{1+C}{C} \left( K_{i}+m\varepsilon
_{i}\right) t^{\frac{C}{1+C}}\right] ,i=1,2,3$. Since generally $C>0$ ,
there is no isotropic limit for $a_{i}$. In the case of small $u$ (very slow
expansion of the universe) from Eq. (21) we obtain $V\sim e^{-\frac{u}{B}}\left( K+Bu\right) ^{\frac{K}{B^{2}}}$
, but in this limiting case the solution of the field equations cannot be
expressed in an analytical form.

The evolution of the mean Hubble factor of the universe is represented in
Fig. 2.

\begin{figure}[h]
\epsfxsize=10cm
\centerline{\epsffile{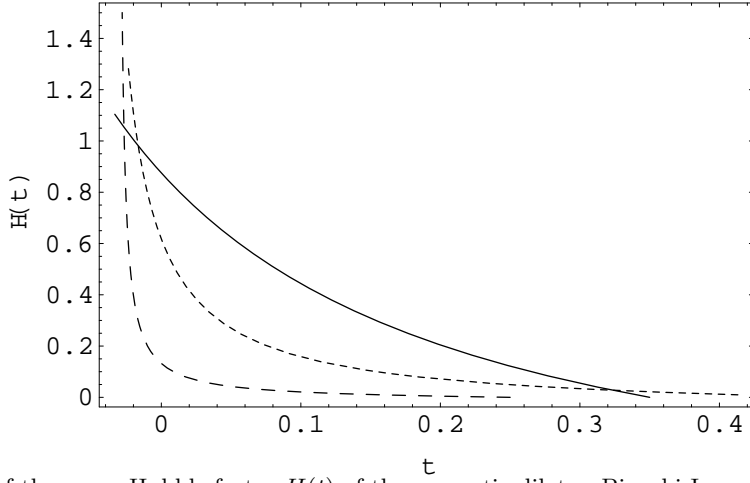}}
\caption{Time evolution of the mean Hubble factor $H(t)$ of the magnetic-dilaton Bianchi I space-time for different
values of the parameters: $B=4$, $C=1$ and $K=1$ ($\Delta >0$) (solid curve), $B=1$, $C=1$, $m=2.5$ and $K=\frac{1}{4}$ ($\Delta =0$) (dotted curve)
and $B=2$, $C=2$ and $K=2$ ($\Delta <0$) (dashed curve).}
\label{FIG2}
\end{figure}

For all three models $H(t)$ is generally a decreasing function of time,
with $H\rightarrow \infty $ for $t\rightarrow 0\ $. In the limit of large $u$
we obtain $H\sim \frac{1}{t}$ .

The dynamics of the mean anisotropy factor, presented in Fig. 3, shows a
rapid time increase of the anisotropy of the dilaton-magnetic field filled
universe.

\begin{figure}[h]
\epsfxsize=10cm
\centerline{\epsffile{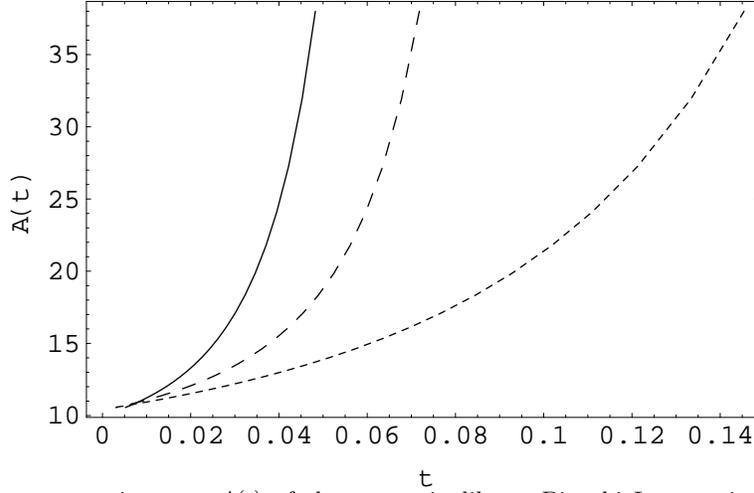}}
\caption{Dynamics of the mean anisotropy $A(t)$ of the magnetic-dilaton Bianchi I space-time for different
values of the parameters: $B=4$, $C=1$ and $K=1$ ($\Delta >0$) (solid curve), $B=1$, $C=1$, $m=2.5$ and $K=\frac{1}{4}$ ($\Delta =0$) (dotted curve)
and $B=2$, $C=2$ and $K=2$ ($\Delta <0$) (dashed curve). The integration constants $K_i,i=1,2,3$ have been normalized so that
$6\sum_{i=1}^{3}\left( \varepsilon _{i}K_{i}+m\right)=1 $ and $3\sum_{i=1}^{3}\left( K_{i}+m\varepsilon _{i}\right) ^{2}=1$.}
\label{FIG3}
\end{figure}

In the limit of large $u$ we have $u=v_{0}t^{-\frac{1}{1+C}%
},v_{0}=const.$ and the mean anisotropy parameter behaves as
\begin{equation}
A=8+2v_{0}\sum_{i=1}^{3}\left( \varepsilon _{i}K_{i}+m\right) t^{\frac{C}{1+C%
}}+3v_{0}^{2}\sum_{i=1}^{3}\left( K_{i}+m\varepsilon _{i}\right) ^{2}t^{%
\frac{2C}{1+C}}. 
\end{equation}

Therefore the inclusion of a magnetic field will increase the anisotropy of
the Bianchi type I space-time.

The dilaton field, presented in Fig. 4, is an increasing function of time. 
In the large $u$ limit it behaves like $\phi \sim \phi _{0}+\frac{\alpha }{2\gamma }\frac{1}{1+C}\ln t+\frac{\alpha m\left( 1+C\right) }{2\gamma C}t^{\frac{C}{1+C}}$
. The dynamics of the universe is non-inflationary, with the deceleration
parameter $q=2+3C$ for $u\rightarrow \infty $.

\begin{figure}
\epsfxsize=10cm
\centerline{\epsffile{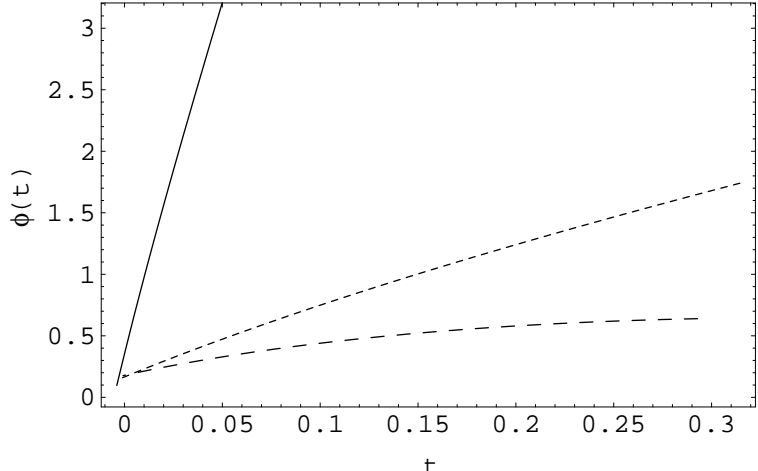}}
\caption{Variation as a function of time of the dilaton field $\phi (t)$ of the magnetic-dilaton Bianchi I space-time for different
values of the parameters: $B=4$, $C=1$ and $K=1$ ($\Delta >0$) (solid curve), $B=1$, $C=1$, $m=2.5$ and $K=\frac{1}{4}$ ($\Delta =0$) (dotted curve)
and $B=2$, $C=2$, $K=2$ and $m=0.0675$ ($\Delta <0$) (dashed curve). The constants $\alpha $, $\gamma $ and $m$ have been normalized so that
$\frac{\alpha }{4C\gamma }=1$ and $\frac{\alpha }{2\gamma C\Delta _{0}}\left( \frac{B}{2C}-m\right) =1$.}
\label{FIG4}
\end{figure}

Finally we shortly discuss the behaviour of the physically important parameters of
the solution in the string frame.

By assuming an anisotropic Bianchi type I geometry with line element $d\hat{s%
}^{2}=d\hat{t}^{2}-\sum_{i=1}^{3}\hat{a}_{i}^{2}\left( \hat{t}\right) \left(
dx^{i}\right) ^{2}$, with the string -frame time coordinate defined
according to $\hat{t}=\int \exp \left( \alpha \phi \right) dt$, the string
frame volume scale factor $\hat{V}$, the directional Hubble factors $\hat{H}%
_{i},i=1,2,3$ and the mean Hubble factor $\hat{H}$ are related by means of
the general transformations \lbrack 12\rbrack
\begin{equation}
\hat{V}=Ve^{3\alpha \phi },\hat{H}_{i}=\left( H_{i}+\alpha \dot{\phi}\right)
e^{-\alpha \phi },i=1,2,3,\hat{H}=\left( H+\alpha \dot{\phi}\right)
e^{-\alpha \phi }. 
\end{equation}

In the string frame the mean anisotropy parameter is given by
\begin{equation}
\hat{A}=\frac{1}{3}\sum_{i=1}^{3}\left( \frac{\hat{H}_{i}-\hat{H}}{\hat{H}}%
\right) ^{2}=\frac{A}{\left( 1+\alpha \frac{\dot{\phi}}{H}\right) ^{2}}.
\end{equation}

With the use of Eq.(13) we obtain
$\frac{\dot{\phi}}{H}=\frac{3\alpha }{2\gamma }+\frac{\alpha m}{2\gamma }%
\frac{1}{HV}=\frac{3\alpha }{2\gamma }+\frac{3\alpha m}{2\gamma }\frac{1}{u}$.
In the limit of small $u$ we obtain
$\hat{A}\rightarrow 0$, while for large $u$ we have $\hat{A}\sim A$.  The
conformal transformation factor $e^{\alpha \phi }$ is given, with the use of
Eq. (14), by $e^{\alpha \phi }=\Phi _{0}V^{\frac{\alpha ^{2}}{2\gamma }}\exp
\left[ \frac{\alpha ^{2}m}{2\gamma }\int \frac{dt}{V}\right] $, where $\Phi =e^{\alpha \phi _0}=const.$. Therefore
for the string frame scale factors we obtain 
\begin{equation}
\hat{a}_{i}=\hat{a}_{i0}V^{\varepsilon _{i}+\frac{\alpha ^{2}}{2\gamma }%
}\exp \left[ \left( K_{i}+m\varepsilon _{i}+\frac{\alpha ^{2}m}{2\gamma }%
\right) \int \frac{dt}{V}\right], i=1,2,3.
\end{equation}

The mathematical form of the string frame scale factors is very similar to
the form of scale factors in the Einstein frame. Therefore  the dynamics of
the dilaton-magnetic Bianchi type I universe has the same general features
in both frames.

In the present paper we have considered the evolution of a dilaton and
magnetic field filled Bianchi type I space-time. The general solution of the
field equations can be expressed in an exact parametric form and, depending
on the numerical values of some constants, three classes of solutions can be
obtained. These solutions describe expanding universes, with
non-inflationary Einstein frame evolutions. In the Einstein frame the
Bianchi type I geometry does not isotropize and there is no
Friedmann-Robertson-Walker limit of these cosmologies. In these models the
dilaton field cannot provide a physical mechanism able to wash out the
initial anisotropies. But such mechanism can be obtained by considering a
matter component in the energy momentum tensor, a cosmological constant or
some quadratic terms in the Lagrangian \cite{Gi99}, \cite{Gi00}, \cite{ChHaMa00a}, \cite{ChHaMa00b}.  
Therefore it is very unlikely that simple pre-big bang models including only
dilaton and magnetic fields can provide a realistic explanation of the
presence of cosmological magnetic fields as remnants of a dilaton-magnetic
era.

\end{document}